\documentclass[sigconf, nonacm, pdfa]{acmart}

\usepackage{algorithmic}
\usepackage{graphicx}
\usepackage{textcomp}
\usepackage{xcolor}
\usepackage{pdflscape}
\usepackage{array}
\usepackage{natbib}
\usepackage{booktabs}
\usepackage{caption}
\usepackage{tabularx}
\usepackage{multirow}
\usepackage{bm}
\usepackage{bbm}
\usepackage{amsmath}
\usepackage{dsfont}
\usepackage{color,soul}
\usepackage{listings}
\usepackage{subcaption}
\usepackage{float}
\usepackage{xspace}
\usepackage{balance}
\usepackage{makecell}
\newcommand{\sys}{SiriusBI\xspace}

\definecolor{codegray}{rgb}{0.5,0.5,0.5}
\definecolor{codegreen}{rgb}{0,0.6,0}
\definecolor{codeblue}{rgb}{0,0,0.6}
\definecolor{codegray}{rgb}{0.5,0.5,0.5}
\definecolor{backcolour}{rgb}{0.95,0.95,0.92}

\newcommand\vldbdoi{10.14778/3750601.3750610}
\newcommand\vldbpages{4860 - 4873}
\newcommand\vldbvolume{18}
\newcommand\vldbissue{12}
\newcommand\vldbyear{2025}
\newcommand\vldbauthors{Jie Jiang, Haining Xie, Siqi Shen, Yu Shen, Zihan Zhang, Meng Lei, Yifeng Zheng, Yang Li, Chunyou Li, Danqing Huang, Yinjun Wu, Wentao Zhang, Bin Cui, Peng Chen}
\newcommand\vldbtitle{SiriusBI} 
\newcommand\vldbavailabilityurl{https://github.com/Tencent-SiriusAI/SiriusBI}
\newcommand\vldbpagestyle{empty} 

\begin{document}
\title{\sys: A Comprehensive LLM-Powered Solution for Data Analytics in Business Intelligence}


\author{
Jie Jiang$^{1}$,
Haining Xie$^{1}$,
Siqi Shen$^{2}$,
Yu Shen$^{1}$,
Zihan Zhang$^{1}$,
Meng Lei$^{1}$,
Yifeng Zheng$^{1}$,
Yang Li$^{1}$, 
Chunyou Li$^{1}$,
Danqing Huang$^{1}$,
Yinjun Wu$^{3}$,
Wentao Zhang$^{2}$, 
Bin Cui$^{3}$,
Peng Chen$^{1}$
}
\affiliation{
{{$^1$}Department of Data Platform, TEG, Tencent Inc.}\\
{{$^2$}Center of Machine Learning Research, Peking University}\\
$^3$School of Computer Science, Peking University\country{}
}
\affiliation{
$^1$\{zeus, hainingxie, willyushen, rylanzhang, garylei, yifengzheng, thomasyngli,  \\chunyouli, daisyqhuang,  felixxfyang, pengchen\}@tencent.com
\\
$^2$\{shensiqi1009, wentao.zhang\}@pku.edu.cn~~~~~
$^3$\{wuyinjun, bin.cui\}@pku.edu.cn\country{}
}

\begin{abstract}

With the proliferation of Large Language Models (LLMs) in Business Intelligence (BI), existing solutions face critical challenges in industrial deployments: functionality deficiencies from legacy systems failing to meet evolving LLM-era user demands, interaction limitations from single-round SQL generation paradigms inadequate for multi-round clarification, and cost for domain adaptation arising from cross-domain methods migration.

We present \sys, a practical LLM-powered BI system addressing the challenges of industrial deployments through three key innovations: 
(a) An end-to-end architecture integrating multi-module coordination to overcome functionality gaps in legacy systems; 
(b) A multi-round dialogue with querying mechanism, consisting of semantic completion, knowledge-guided clarification, and proactive querying processes, to resolve interaction constraints in SQL generation; 
(c) A data-conditioned SQL generation method selection strategy that supports both an efficient one-step Fine-Tuning approach and a two-step method leveraging Semantic Intermediate Representation for low-cost cross-domain applications. Experiments on both real-world datasets and public benchmarks demonstrate the effectiveness of \sys. User studies further confirm that \sys enhances both productivity and user experience.

As an independent service on Tencent's data platform, \sys is deployed across finance, advertising, and cloud sectors, serving dozens of enterprise clients. It achieves over 93\% accuracy in SQL generation and reduces data analysts' query time from minutes to seconds in real-world applications.
\end{abstract}

\maketitle

\pagestyle{\vldbpagestyle}
\begingroup\small\noindent\raggedright\textbf{PVLDB Reference Format:}\\
\vldbauthors. \vldbtitle. PVLDB, \vldbvolume(\vldbissue): \vldbpages, \vldbyear.\\
\href{https://doi.org/\vldbdoi}{doi:\vldbdoi}
\endgroup
\begingroup
\renewcommand\thefootnote{}\footnote{\noindent
This work is licensed under the Creative Commons BY-NC-ND 4.0 International License. Visit \url{https://creativecommons.org/licenses/by-nc-nd/4.0/} to view a copy of this license. For any use beyond those covered by this license, obtain permission by emailing \href{mailto:info@vldb.org}{info@vldb.org}. Copyright is held by the owner/author(s). Publication rights licensed to the VLDB Endowment. \\
\raggedright Proceedings of the VLDB Endowment, Vol. \vldbvolume, No. \vldbissue\ %
ISSN 2150-8097. \\
\href{https://doi.org/\vldbdoi}{doi:\vldbdoi} \\
}\addtocounter{footnote}{-1}\endgroup

\ifdefempty{\vldbavailabilityurl}{}{
\vspace{.3cm}
\begingroup\small\noindent\raggedright\textbf{PVLDB Artifact Availability:}\\
The source code, data, and/or other artifacts have been made available at \url{\vldbavailabilityurl}.
\endgroup
}

\section{Introduction}

Business Intelligence (BI)~\cite{DBLP:journals/cais/Negash04/BIcite1,DBLP:conf/centeris/WiederO15/BIcite2} is a crucial application scenario in the data field, comprising a comprehensive suite of methodologies, tools, and infrastructures designed to collect, integrate, analyze, and present raw data from an organization to generate actionable insights for informed decision-making. 
BI systems are extensively used in various sectors, including finance~\cite{nuseir2021designing/X24f-apply}, environment~\cite{DBLP:conf/ifip5-7/HeggerS24-envir}, and social media~\cite{DBLP:journals/jcmse/ChenWW24a-apply,DBLP:conf/gamesem/Rivera24-media}, which significantly improves the decision-making process through the provision of real-time analytics and reporting capabilities~\cite{wang2025largedse, li2024opengaussjcst}.

A typical BI system comprises several key components: a data management module that stores, processes, and aggregates vast amounts of data; analytic algorithms that transform the data into actionable insights; and visualization tools that present the information in intuitive and user-friendly formats.
Among these, data analytics plays a crucial role in providing decision-making support, directly determining the correctness and appropriateness of decisions.
Recent advancements in LLMs~\cite{DBLP:journals/corr/abs-2405-00527/chatbi, ke2024unveilingscis, zhang2024spikingminilmscis} have sparked significant interest in ChatBI — a new paradigm supported by natural language interfaces~\cite{nlisurvey}. 
Concurrently, the demand for a fully integrated and efficient ChatBI solution is surging, driven by the need of a more intuitive and accessible mode of data interaction. 
This evolution promises to transform how users engage with data, making insights more available and actionable.

\begin{figure*}
  \centering
  \includegraphics[width=0.95\textwidth]{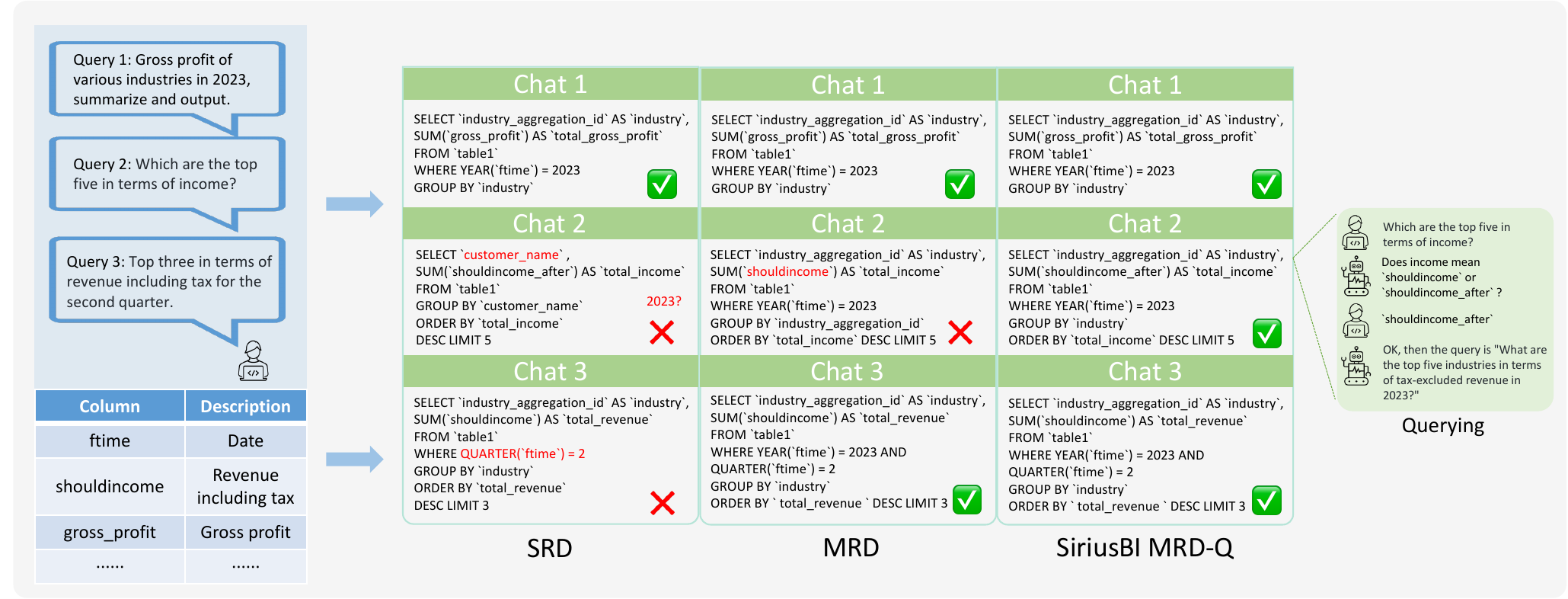}
  \vspace{-10pt}
  \caption{Demonstration of multi-round user requests. Compared with \sys, SRD misses the omitted year information in conversations, while MRD fails to identify the user's ambiguous intent.}
  \label{fig:MRD-Qshowcase}
  \vspace{-10pt}
\end{figure*}

To meet the growing demand for big data analytics and decision-making in BI, the data community has proposed numerous effective approaches. However, when applying existing work in real-world BI scenarios, we identify the following three challenges: 

\textbf{C1: Functionality Deficiencies}. 
While traditional business intelligence systems~\cite{chaudhuri2011overviewBI} integrate core components spanning data management, SQL generation, and insight discovery to form complete analytics pipelines, their reliance on heuristic rules and conventional AI/ML techniques limits generalization ability in dynamic scenarios. Although LLM-based methods have advanced task-specific performance, few offer comprehensive BI capabilities comparable to their traditional counterparts. 
For example, MAC-SQL~\cite{DBLP:journals/corr/abs-2312-11242/macsql} and CHESS~\cite{chesssql} optimize NL2SQL accuracy but treat SQL execution as terminal outputs, neglecting downstream tasks like attribution analysis. While Lian et al.~\cite{DBLP:journals/corr/abs-2405-00527/chatbi} extend their pipeline with Apache Superset for visualization, they fail to introduce knowledge bases to support dynamic grounding of domain-specific context, a critical requirement for real-world BI adaptation~\cite{cheng2009ontology}. This functional fragmentation forces users to manually coordinate tools (e.g., SQL editors, dashboard platforms, knowledge retriever), which imposes significant cognitive load and reduces operational efficiency, as evidenced by industry report~\cite{gartner2024}.

\textbf{C2: Interaction Limitations}. 
In the context of ChatBI, the NL2SQL task is becoming increasingly vital, as it facilitates seamless interaction between natural language queries and structured data retrieval, thereby enhancing the efficiency and accuracy of data analytics. The evolution of NL2SQL techniques reveals a critical architectural mismatch: while traditional methods (schema-based~\cite{HristidisP02,LuoLWZ07,ZengLL16,HristidisGP03} or parsing-based~\cite{IyerKCKZ17,WangCB17,LiJ14,DBLP:conf/coling/PopescuAEKY04}) and modern LLM-driven approaches (prompt engineering~\cite{DBLP:conf/nips/PourrezaR23/dinsql,DBLP:journals/corr/abs-2312-11242/macsql} or fine-tuning techniques~\cite{li2024codes,DBLP:journals/corr/abs-2402-01117/dts-sql}) predominantly optimize for \textbf{S}ingle-\textbf{R}ound \textbf{D}ialogue (\textbf{SRD}) precision. This SRD-centric paradigm introduces a significant continuity gap in \textbf{M}ulti-\textbf{R}ound \textbf{D}ialogues (\textbf{MRD}): real-world BI workflows often require iterative investigation through successive queries, where later queries tend to omit previously provided contextual information, resulting in semantic ambiguity beyond the initial query. For instance, in the MRD NL2SQL task illustrated in Figure~\ref{fig:MRD-Qshowcase}, the user issues three queries; notably, the second and third queries omit the time condition ``2023'' because it was specified in the first query. 
Single-round NL2SQL approaches demand nearly perfect input specificity, which explains its failure to generate correct SQL statements for the second and third queries in Figure~\ref{fig:MRD-Qshowcase}.
Worse still, due to the intricate nature~\cite{NL2SQL-survey} of MRD, few approach has been devoted to addressing this task.
Lian et al.'s MRD solution~\cite{DBLP:journals/corr/abs-2405-00527/chatbi} is the first attempt towards this task. Nevertheless, their solution is absent of user-guided clarification loops for intent resolution and domain-grounded dialogue act modeling. As demonstrated in Figure~\ref{fig:MRD-Qshowcase}, the basic MRD approach still exhibit performance degradation beyond the first dialogue round.

\textbf{C3: Cost for Domain Adaptation}.
Cross-domain deployment of NL2SQL models faces the challenge of cost surges in domain knowledge transfer, primarily caused by insufficient model generalization capability.
Structural differences in database schema across domains (e.g., nested tables in finance vs. wide tables in advertising) necessitate repetitive model adaptation~\cite{NL2SQL-survey}, while semantic gaps between industry-specific operators (e.g., financial window functions vs. e-commerce promotional rules) exacerbate logical deviations in SQL generation~\cite{LiLCLT24/nl2sql360}. Critically, domain knowledge transfer relies heavily on expert-annotated data, with manual annotation costs growing with domain complexity~\cite{sea-sql}. Our real-world deployment statistics show that direct model migration leads to business logic errors in approximately two-thirds of generated SQL queries. Meanwhile, adapting models through traditional fine-tuning requires 5.5 person-days on average to label 200 seed queries within existing databases—forming critical bottlenecks for enterprise-level scalability. 

To address the aforementioned challenges, we propose \sys, which implements a comprehensive LLM-powered solution for ChatBI scenarios. 
This system leverages the capabilities of LLMs to empower various modules, thereby enhancing both the efficiency and user experience in data analytics. 
Specifically, for the issue of functionality deficiencies (\textbf{C1}), \sys introduces an \textbf{end-to-end} integrated architecture that seamlessly orchestrate core modules including knowledge management, multi-round dialogue analysis, SQL generation, and data insight provision, thereby ensuring a closed-loop pipeline from natural language queries to final decision-making reports.

For the issue of interaction limitations (\textbf{C2}), we introduce the \textbf{MRD-Q} (\textbf{M}ulti-\textbf{R}ound \textbf{D}ialogue with \textbf{Q}uerying) module. 
As a supplement to the basic multi-round dialogue analysis module proposed by Lian et al.~\cite{DBLP:journals/corr/abs-2405-00527/chatbi}, MRD-Q incorporates an intent querying module to clarify user queries through follow-up questions. 
This approach enables the system to accurately identify the user's true intent, even when the initial query is incomplete or ambiguous, thus facilitating precise responses, as shown in Figure~\ref{fig:MRD-Qshowcase}. 

\begin{figure*}[t]
  \centering
  \includegraphics[width=0.95\textwidth]{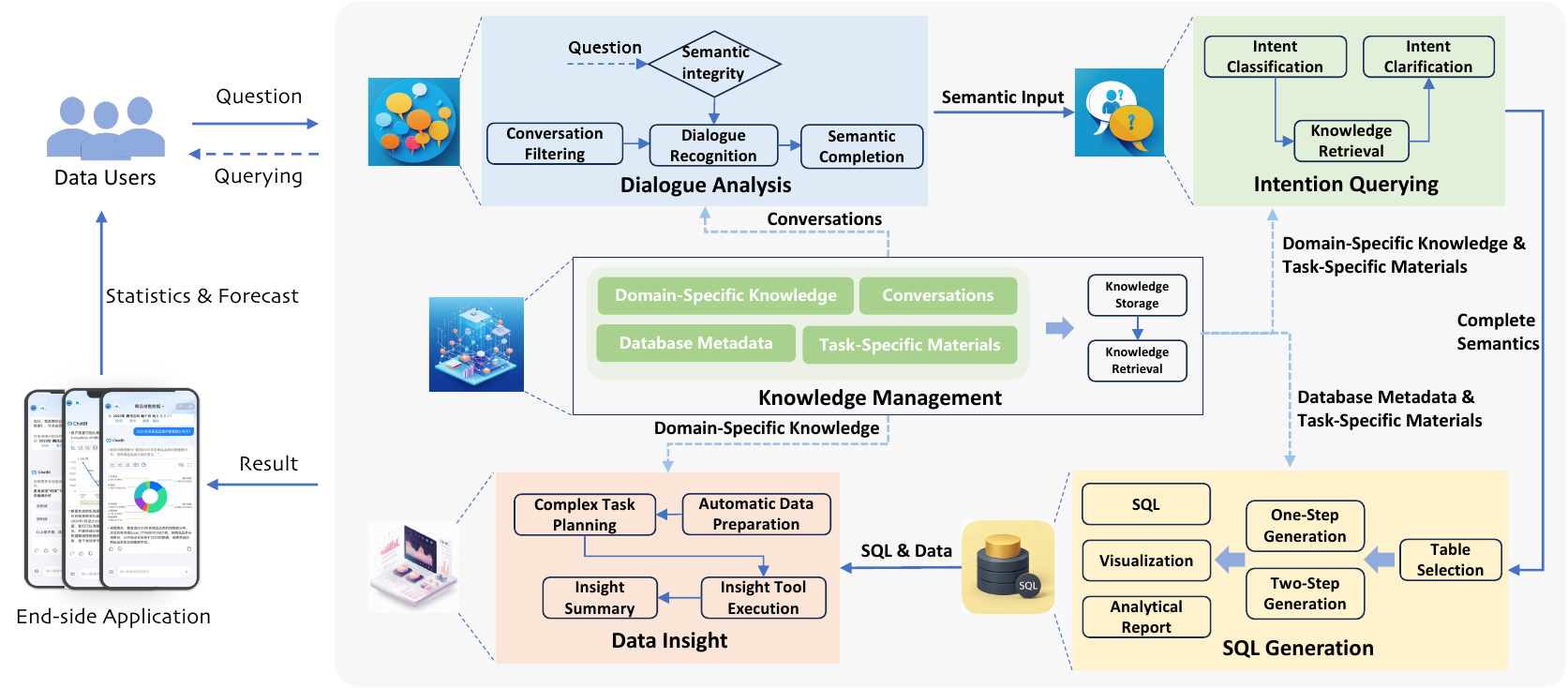}
  \vspace{-10pt}
  \caption{Workflow overview of \sys.}
  \label{fig:SiriusBI}
  \vspace{-10pt} 
\end{figure*}

To enable economic domain adaptation (\textbf{C3}), \sys introduces a strategy switching mechanism that dynamically selects between one-step and two-step SQL generation paradigms based on data conditions. This mechanism optimizes the trade-off between computational cost and performance, ensuring efficient adaptation to diverse business domains.
To address structural and semantic disparities across domains, the one-step method employs a supervised fine-tuning (SFT) method enhanced by an automated data generation process pipeline. This pipeline minimizes manual annotation costs by leveraging SQL logs and LLM-based reverse engineering to synthesize high-quality training data, greatly reducing reliance on labor-intensive expert labeling. For cold-start scenarios, the two-step method eliminates SFT by combining domain knowledge retrieval with semantic intermediate representation (SIR), achieving robust performance across diverse industries without domain-specific tuning. 
The generalization ability of our system is further enhanced by a knowledge base with hybrid storage strategies, specially designed for BI application scenario.

\textbf{Contributions.} The main contributions of this paper can be summarized as follows:
\begin{itemize}
    \item \underline{Practical ChatBI System}: 
    \sys integrates four core functionalities — knowledge management, multi-round dialogue with querying, SQL generation, and data insight — into a unified system. It has been deployed across Tencent's finance, advertising, and cloud businesses departments, serving dozens of enterprise clients, reducing data analysis latency from minutes to seconds. 
    
    \item \underline{New Dialogue Analysis Mechanism}: 
     We propose a new dialogue analysis mechanism named MRD-Q, consisting of semantic completion, knowledge-guided clarification, and proactive querying processes. It effectively addresses issues raised by ambiguous or incomplete queries in real-world scenarios.

    \item \underline{Economic Domain Adaptation Strategy}: We propose a dynamic strategy selection mechanism for SQL generation conditioned on data characteristics. To reduce annotation costs for fine-tuning, we set up an automated data synthesis pipeline. To enhance cross-domain SQL syntax accuracy and business logic consistency, we use Semantic Intermediate Representation (SIR) in prompt engineering.
    
    \item \underline{Multi-Round NL2SQL Benchmark}: We construct \textit{MRD-BIRD} —a dataset containing 96 multi-round dialogue sequences (3–5 interaction rounds)— and open-source the data to advance NL2SQL research in multi-round dialogue contexts.
    
    \item \underline{Proven Effectiveness}: Extensive experiments on both academic benchmarks and real-world datasets demonstrate the effectiveness of \sys.
    Notably, when applied in industry, \sys achieves remarkable accuracy rates of \textbf{97\%} in SQL generation for Tencent Finance, \textbf{93\%} for Tencent Advertisement, and \textbf{96\%} for Tencent Cloud. User studies further confirm that \sys enhances both productivity ad user experience.
\end{itemize}

\section{Related Work}

\subsection{Business Intelligence}
The integration of AI and Machine Learning (ML) has transformed data-driven decision-making in BI~\cite{rane2024business/relat-1}. ML and Predictive Analytics enhance forecasting and pattern recognition in large datasets~\cite{lee2022predictive/relat-2,tamang2021improving/relat-3}. 
NLP and Text Analytics extract insights from unstructured data, improving customer service and decision-making~\cite{kaur2019case/relat-4,ittoo2016text/relat-5}. 
The rise of the ChatBI paradigm, which has significantly advanced business intelligence and taken a major step towards data democratization~\cite{linkon2024advancements/relat-10}, is largely driven by the development of advanced AI models such as BERT~\cite{devlin2019bert} and ChatGPT~\cite{OpenAI}, which enable intelligent data interpretation and improved user interactions. Among related works, BI-REC~\cite{meduri2021birec} is a pioneering system that recommends BI applications to help users achieve their data analysis goals. Similar to our method, BI-REC includes a step for predicting BI intents. However, BI-REC focuses on leveraging historical state information to predict possible future BI intents, whereas our approach aims to assess whether the current user intent is sufficiently clear and precise to generate an accurate SQL query.
Lian et al.~\cite{DBLP:journals/corr/abs-2405-00527/chatbi} are the first to recognize the importance of the NL2BI task in practical ChatBI applications and propose an efficient method to address it. We draw inspiration from their dialogue analysis technique to obtain semantically complete queries. The key differences between our approach and theirs are: (1) we leverage LLMs instead of smaller pre-trained models to analyze query completeness; (2) we incorporate domain-specific knowledge to help the LLM better understand user intentions; and (3) we enable active clarification questions directed at users when the LLM determines that the query requires further disambiguation. Lian et al.'s method does not consider the latter two aspects.

\subsection{NL2SQL Methods}
NL2SQL is a core task in ChatBI systems and can be broadly categorized into prompt-based and training-based methods~\cite{zhou2024dbdse}. Recently, Large Language Models (LLMs) like Codex~\cite{codex} and Claude 3~\cite{claude3} have shown great promise in NL2SQL, leveraging techniques such as prompt engineering~\cite{giray2023prompt/prompteng}, in-context learning~\cite{dong2022survey/in-context}, and chain-of-thought prompting~\cite{wei2022chain/cot}. Prompt-based approaches not only offer new technical paradigms but also reduce costs in industrial settings. Notable examples include DIN-SQL~\cite{DBLP:conf/nips/PourrezaR23/dinsql}, C3~\cite{dong2023c3/c3}, MAC-SQL~\cite{DBLP:journals/corr/abs-2312-11242/macsql}, and SQL-PaLM~\cite{sun2023sql/sql-palm}, which improve NL2SQL performance on models like ChatGPT and PaLM~\cite{narang2022pathways/palm}.

Before LLMs, NL2SQL solutions mainly relied on training encoder-decoder sequence-to-sequence models~\cite{cai2021sadga/sql1,cao2021lgesql/sql2,choi2021ryansql/sql3,gan2021natural/sql4,wang2020rat/sql5}, requiring large amounts of database and SQL data. The advent of LLMs has introduced new training paradigms; for instance, Codes~\cite{li2024codes} proposes a training method tailored for NL2SQL that boosts LLM performance. However, some studies~\cite{maamari2024death,pourreza2024/chase,caferouglu2024sql/esql} fine-tune LLMs like ChatGPT~\cite{OpenAI} and Gemini~\cite{gemini} to further improve results, though this often incurs high practical costs.
\section{Method}

\subsection{Overview}
The overall framework of \sys is illustrated in Figure~\ref{fig:SiriusBI} and consists of four core modules:

\noindent(1) The \textbf{Knowledge Management} module consists of four types of data -- database metadata, domain-specific knowledge, task-related materials (e.g., demonstrations used for in-context learning in LLMs), and historical conversations. These data types are stored in a hybrid architecture, ensuring efficient and scalable access to knowledge, supporting the seamless functionality of other modules.

\noindent(2) The \textbf{Multi-Round Dialogue with Querying (MRD-Q)} module is composed of two components: \textit{Dialogue Analysis} and \textit{Intention Querying}, which enable multi-turn user-oriented dialogue and knowledge-guided clarification to resolve semantic ambiguity in Conversational BI. Unlike existing SRD-centric approaches~\cite{DBLP:journals/corr/abs-2312-11242/macsql,DBLP:conf/nips/PourrezaR23/dinsql}, MRD-Q introduces follow-up question generation based on domain-specific knowledge, enabling context-aware SQL calibration across 3-5 dialogue turns.

\noindent(3) The \textbf{SQL Generation} module dynamically selects between one-step (fine-tuning based) and two-step (semantic intermediate representation based) processing paradigms according to business scenario data characteristics. This dual-mode approach efficiently transforms user intent into SQL queries with optimal cost-effectiveness.
The one-step method streamlines the process through an automated data preparation pipeline, while the two-step method employs a knowledge-base-driven query rewriting strategy that eliminates model training requirements. This architectural design balances operational efficiency with implementation flexibility across diverse application contexts.

\begin{figure}[htbp!]
    \centering
    \includegraphics[width=\linewidth]{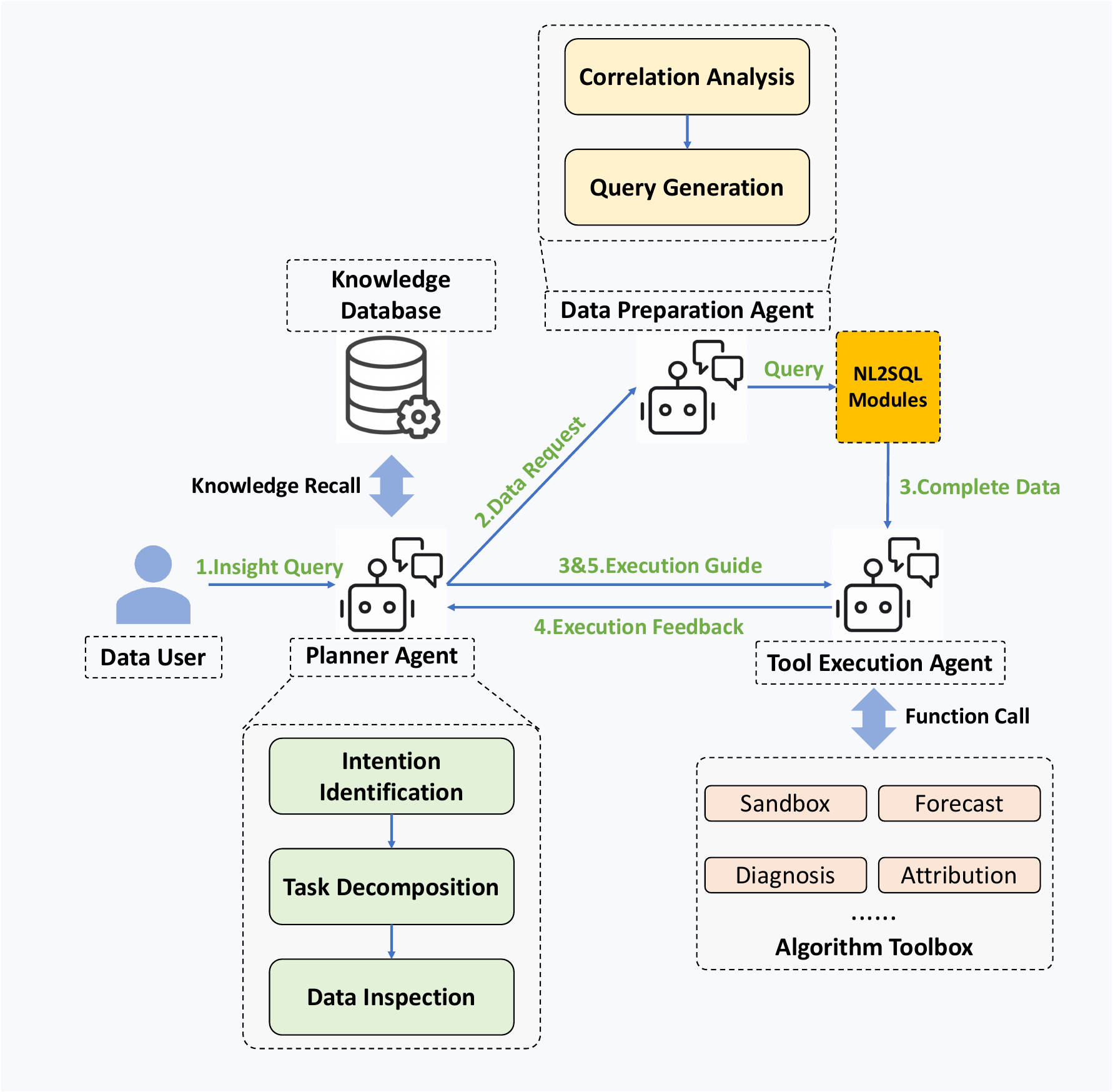}
    \vspace{-15pt}
    \caption{Detailed description of the task planning workflow for \sys data insights in response to user queries.}
    \label{fig:data_insight_workflow}
    \vspace{-10pt}
\end{figure}

\noindent(4) To close the loop from SQL execution to decision support, \sys integrates a \textbf{Data Insight} module that interprets query results through a multi-agent workflow based on the FunctionCall ~\cite{kimllm/functioncall} and ReAct~\cite{react} mechanism as shown in Figure ~\ref{fig:data_insight_workflow}. 
Specifically,   
the \textit{Planner Agent} analyzes user queries, SQL results, and domain knowledge to decompose tasks into subtasks and issue instructions;
the \textit{Data Preparation Agent} identifies missing data requirements and generates SQL queries to retrieve necessary information; 
the \textit{Tool Execution Agent} selects specialized tools (Forecast/Diagnosis/Attribution Tools, implemented via reproduced existing works ~\cite{forecasttool,detecting,attribution-adtributor,attribution-generic,attribution-unified}) to process data, with final outputs consolidated by the Planner Agent for response generation.

\begin{figure}[htbp!]
  \centering
  \includegraphics[width=0.9\linewidth]{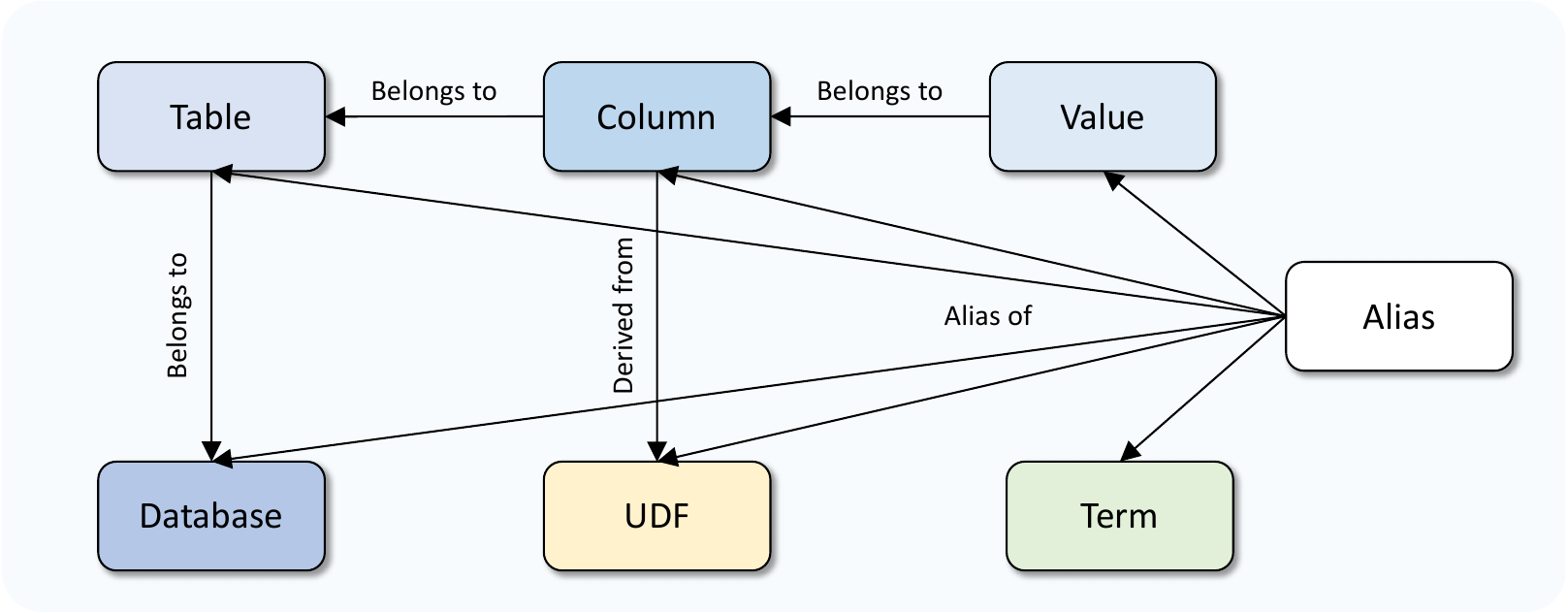}
  \caption{The hierarchical structure of database schema and some related domain-specific knowledge.}
  \label{fig:topological_kw}
  \vspace{-10pt}
\end{figure}

\begin{figure*}[htbp!]
  \centering
  \includegraphics[width=\linewidth]{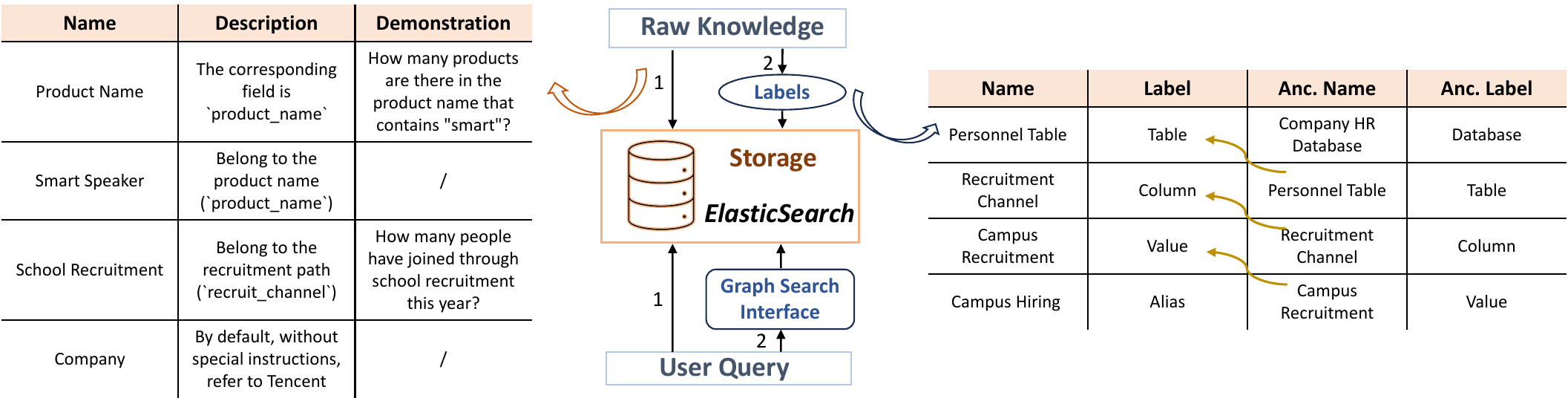}
  \vspace{-15pt}
  \caption{Illustration of the hybrid storage strategy.}
  \label{fig:knowledge_management}
  \vspace{-10pt}
\end{figure*}

\subsection{Knowledge Management}
\label{sec_km}

The application of BI often requires substantial amount of domain-specific knowledge. To address the knowledge requirements, \sys constructs and maintains a comprehensive knowledge base within the knowledge management module. This knowledge base is designed to support efficient data retrieval, enhance decision-making processes, and facilitate seamless integration with BI tools.

The knowledge base primarily constitutes four types of data:
(1) \underline{Database Metadata}: This includes essential information such as table names, column types, column names, and other details of the database schema.
(2) \underline{Domain-Specific Knowledge}: This encompasses explanations of certain fields, term definitions, and relevant business context. It provides the necessary background information to interpret and utilize the data effectively within the specific domain.
(3) \underline{Task-Specific Materials}: These are groups of prompts, demonstrations and other resources designed for different tasks. They are tailored to guide users in performing specific operations or analyses within the BI system.
(4) \underline{Historical Conversations}: This includes logs of past interactions and queries, which can be used to complete semantic for multi-round dialogue analysis.

\subsubsection{Knowledge Storage}
We observe that certain types of knowledge entries exhibit dependencies, while others remain relatively independent. In our application scenario, the relationships between database metadata and relevant domain-specific knowledge can be represented as a directed acyclic graph as illustrated in Figure~\ref{fig:topological_kw}. To ensure comprehensive retrieval of all related knowledge, it is essential to explicitly capture and represent these relationships. 

While graph databases~\cite{easygraph} offer a natural solution for storing such relational knowledge, we opt against their use due to their inherent complexity and overhead, which exceed the requirements of our system. Instead, we adopt a lightweight approach by augmenting knowledge entries with two dedicated fields: \texttt{Anc. Label}, which specifies the label of the ancestor entry, and \texttt{Anc. Name}, which encodes the concrete value of the ancestor entry. This design, as depicted in Figure~\ref{fig:knowledge_management} (right table), preserves the relational semantics necessary for knowledge retrieval while avoiding the operational and computational complexities associated with native graph storage. This approach strikes a balance between functionality and efficiency, aligning with the practical needs of our system.

For independent knowledge entries, we simply store a description of the entry and, if applicable, a demonstration or example. This straightforward approach is sufficient for entries that do not have dependencies or relationships with other knowledge items, as shown in Figure~\ref{fig:knowledge_management} (left table).


\subsubsection{Knowledge Extraction} 
\label{sec_kr}
The objective of the knowledge extraction process in \sys is to efficiently and accurately extract relevant information from the knowledge base. We employ different knowledge extraction strategies based on the storage strategy described earlier. All queries must undergo at least one round of knowledge retrieval, which is divided into two main phases: coarse retrieval and fine retrieval. This two-phase approach ensures a balance between recall and precision, enabling the system to first identify a broad set of potentially relevant knowledge and then refine the results to meet the specific needs of the task at hand.

In the coarse retrieval phase, the primary goal is to maximize recall, ensuring that a comprehensive set of potentially relevant knowledge entries is retrieved. This is particularly important for tasks such as SQL generation, where the system needs to identify all possible column names or metadata that might be relevant to the user's query. To achieve this, we employ a mixed-metric search strategy that operates on both lexical and semantic levels~\cite{wang2024multilingual}. Lexical search captures exact matches or close variants of the query terms, while semantic search leverages embeddings to identify conceptually related knowledge entries, even if they do not share exact lexical overlap. For this phase, we utilize Elasticsearch~\cite{elasticsearch2018elasticsearch}, a widely adopted search engine known for its scalability and efficiency. The query is processed without segmentation, allowing for a broader initial recall set.

For knowledge entries that exhibit hierarchical relationships (e.g., those with \texttt{Anc. Name} fields), we implement a multi-round retrieval strategy inspired by multi-hop graph traversal. Specifically, for each entry retrieved in the initial search, if its \texttt{Anc. Name} field is non-empty, we recursively retrieve the ancestor entries. This process continues iteratively until no further ancestor entries are found. This approach ensures that the retrieval process captures not only the directly relevant knowledge but also the contextual and hierarchical relationships that are critical for tasks such as SQL generation or business rule interpretation. 

To illustrate this process, consider the right table in Figure~\ref{fig:topological_kw}. The retrieval begins with the entry ``Campus Hiring'', whose \texttt{Anc. Name} field contains the ancestor entry ``Campus Recruitment''. The system then retrieves ``Campus Recruitment'', whose \texttt{Anc. Name} field includes ``Recruitment Channels''. Next, ``Recruitment Channels'' is retrieved, which further links to ``Personnel Table''. Finally, ``Personnel Table'' is retrieved, whose \texttt{Anc. Name} field points to ``Company HR Database''. Since ``Company HR Database'' is not present in the knowledge base, the retrieval process terminates. This example demonstrates how the multi-hop traversal strategy effectively captures a chain of related knowledge entries, ensuring comprehensive retrieval of contextually relevant information.

The coarse retrieval phase typically yields a large set of candidate knowledge entries. To refine these results, we perform a fine retrieval phase, which focuses on improving precision by re-ranking and filtering the candidate set. We employ a state-of-the-art re-ranking model, BGE~\cite{chen2024bge}, to rank the candidate nodes based on their relevance to the query. 
For tasks such as SQL generation, we further refine the results by leveraging a LLM to filter out irrelevant columns. The LLM evaluates the semantic relevance of each column to the user query and removes those that do not contribute to the intended SQL query. This step ensures that the final set of retrieved knowledge is both concise and highly relevant to the task.

By combining coarse retrieval, relationship-aware multi-hop traversal, re-ranking, and LLM-based filtering, \sys achieves a robust and efficient knowledge retrieval process. This approach not only ensures high recall and precision but also adapts to the complex dependencies and relationships inherent in domain-specific knowledge, making it well-suited for a wide range of BI applications.

\subsection{Multi-Round Dialogue with Querying}
\label{sec_MRDQ}
\textbf{M}ulti-\textbf{R}ound \textbf{D}ialogue with \textbf{Q}uerying (\textbf{MRD-Q}) is a critical component in \sys that receives and process user inputs for data integrity.
It is designed to support MRD through automatic semantic completion and user-involved intent querying. A detailed introduction is presented in the following sections.

\subsubsection{Dialogue Analysis}
\label{sec_dia_ana}

The multi-round dialogue analysis mechanism in \sys is built upon the basic dialogue analysis module in previous work\cite{DBLP:journals/corr/abs-2405-00527/chatbi}. 
It first determines the semantic integrity of the user query by assessing whether the current input or historical conversations contain \textit{metrics} and \textit{dimensions}. 
In the context of business intelligence, dimensions refer to attributes that provide granularity level information of the user query, such as time periods or geographical locations. For example, ``last quarter'' is a time dimension and ``in New York'' is of location dimension. Metrics, on the other hand, are quantitative measurements, such as the number of sales, which might be represented by a column like \texttt{Total\_Sales}. When a query includes both dimensions and metrics, such as ``the number of sales in New York last month'', it is possible to construct a corresponding SQL query. Conversely, if a query lacks either a dimension or a metric, such as ``How about Los Angeles?'', it becomes challenging to generate a meaningful SQL query. Although this approach is somewhat intuitive, it effectively addresses most dialogue patterns encountered in BI scenarios.

If both are found, the current input is considered of semantic integrity, and the workflow proceeds to the next module. 
If either is missing, the system supplements the current input's semantics by selecting the most recent historical input that includes metrics and dimensions, and then moves to the next step.

Unlike previous work, \sys does not fine-tune a PLM (Pre-trained Language Model, e.g., ~\cite{DBLP:journals/corr/abs-2405-00527/chatbi} adopts ERNIE~\cite{sun2019ernie}) to determine the semantic integrity or to select relevant historical conversations. PLMs, such as BERT~\cite{devlin2019bert}, GPT-2~\cite{radford2019language}, and ERNIE~\cite{sun2019ernie}, are large-scale neural networks pre-trained on vast amounts of text data to capture general language representations. While they have shown remarkable performance in various natural language processing tasks, fine-tuning them for specific applications often demands substantial labeled data and computational resources. Additionally, fine-tuned models may not generalize well across different domains, limiting their transferability and effectiveness in diverse BI scenarios. Instead, we leverage the robust reasoning capability of LLMs to figure out the existence of ``metric'' and ``dimension'' and the historical conversations. 

\subsubsection{Intention Querying}

For multi-round dialogues in real-world scenarios, relying solely on the dialogue history is often insufficient to complete the missing information due to the inherent complexity. To address these challenges, we enhance the basic dialogue analysis module by introducing an intent querying feature, resulting in a robust MRD-Q component. The intention querying process consists of the following steps:

\textit{Intent Classification}.
We use the same LLM for dialogue analysis in Section~\ref{sec_dia_ana} and further categorize the expanded queries into three types. Denote the user input as $Q$, and the result of the intent classification as $I(Q)$. Formally, we classify the query into three categories, as indicated by Eqn.~\eqref{eq:intent_classification}.
\begin{equation}
I(Q) = 
\begin{cases} 
2, & \textit{if }\, Q \textit{ contains both dimensions and columns}, \\
1, & \textit{if }\, Q \textit{ lacks either dimensions or columns}, \\
0, & \textit{if }\, Q \textit{ is a non-BI scenario question}.
\end{cases}
\label{eq:intent_classification}
\end{equation}
When $I(Q)=0$, the query is identified as out of the BI scope. In this case, the system will politely decline the request and guide the user to ask data analysis-related questions. The dialogue analysis process will then restart.
When $I(Q)=1$, it indicates that, even after supplementing with historical inputs, the user query still lacks key information. The system will proactively ask the user to provide the missing information and then restart the analysis process.
When $I(Q)=2$, the query is complete. The system will proceed to the next step of the pipeline.



\textit{Intent Clarification.}
Although the queries entering this stage are semantically complete, additional processing is required to ensure the generated SQL accurately reflects the user’s intent due to real-world complexities. To address this, the module first retrieves domain-specific knowledge from the knowledge base based on the user's query. For instance, the triplet associated with the term ``YTD'' is defined as: (``YTD'', ``Year To Date or Year To Day, referring to the period from the beginning of the current year to the present date.'', ``What is the YTD revenue of the mini-program?''). Such knowledge enhances the LLM’s understanding of domain-specific terminology. Furthermore, the module retrieves task-specific materials, including prompt templates and demonstration examples, to guide the LLM in intent clarification. Specifically, the LLM either returns a semantically refined query or interactively presents users with disambiguation options to ensure input precision.


In summary, we develop the MRD-Q component, which supplements the heuristic dialogue analysis by the intent clarification capability of the LLM. Even if the user provides arbitrary answers, \sys can clarify the user's true intent through guided follow-up questions, therefore improving the successful rate of the task execution.

\subsection{SQL Generation}
\label{sec_sql_gen}
NL2SQL is the core task within the entire BI system, and the performance bottleneck continues to pose challenges in the industry. 
In this subsection, we will focus on the implementation approach of the SQL Generation module in \sys, including table selection, as well as the SQL generation methods. 

\subsubsection{Table Selection}
Before generating SQL queries, it is necessary to select the appropriate tables from numerous options in an industrial database based on the user's input.
This process relies on retrieving the relevant table information mentioned in the user's query. 
The implementation of table selection is divided into two parts: coarse ranking and re-ranking. 
During the coarse-ranking stage, the number of candidate tables is typically reduced by a factor of ten or more to maintain fewer than 100 candidates, thereby ensuring computational efficiency in the subsequent fine-ranking phase.
In the re-ranking stage, these candidates are re-ranked to identify the most relevant ones for SQL generation. 
\begin{figure}
  \centering
  \includegraphics[width=\linewidth]{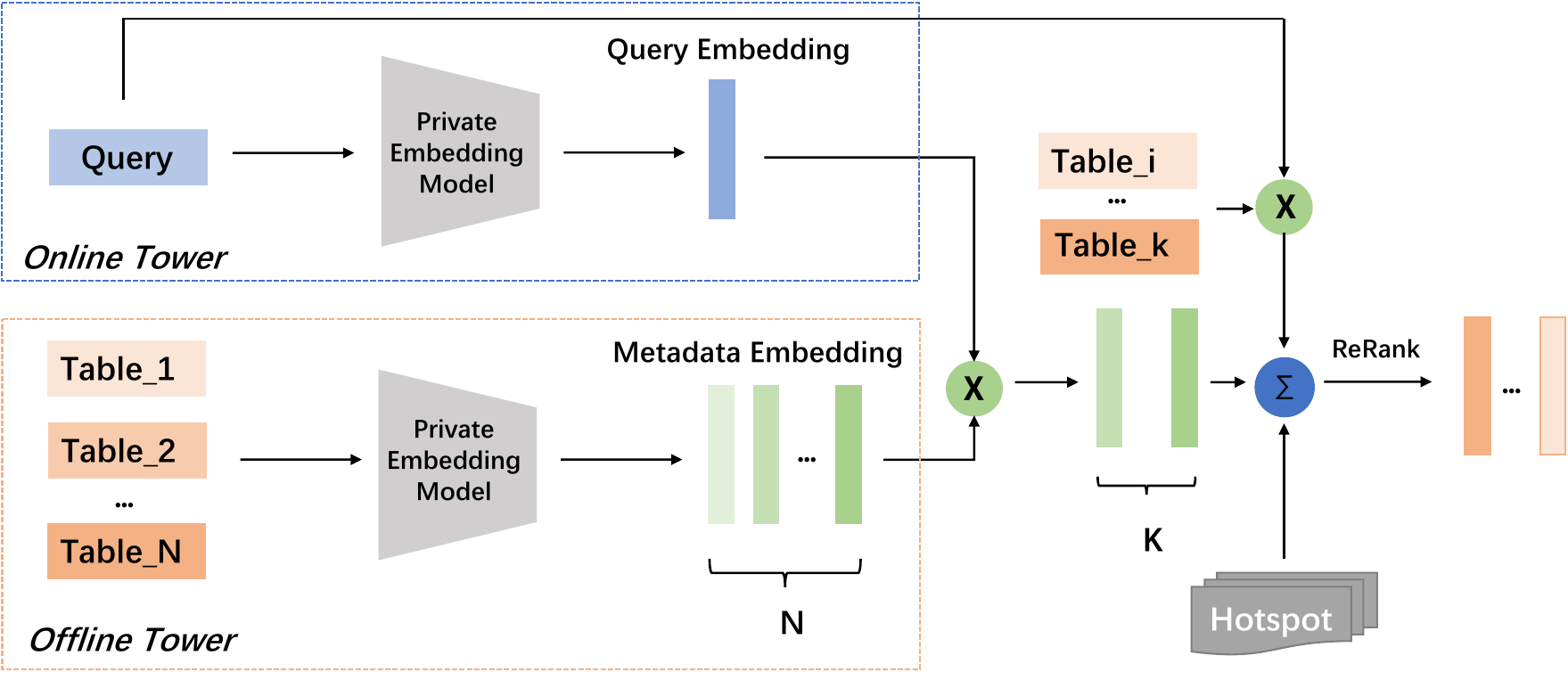}
  \vspace{-15pt}
  \caption{Overview of the table selection process.}
  \label{fig:table selection}
  \vspace{-15pt}
\end{figure}

The coarse ranking process consists of two parts: online and offline. 
In the offline part, each data table is subjected to a private embedding model and transformed into an embedding vector.
We use the StarRocks (SR)~\cite{starrocks} vector database to store the embeddings of tables to ensure data security 
and the embeddings are extracted using the general-purpose pre-trained model m3e-large~\cite{m3e}, ensuring high-quality feature representations and domain adaptability.
SR supports millisecond-level latency retrieval and allows for filtering by specified tags, making it a robust solution for efficient table selection.
In the online part, we extract multiple keywords from the query through a LLM and then perform individual retrieval using each keyword.
Compared to directly embedding the user's input, this approach eliminates redundant information, reduces interference from irrelevant information, and increases the recall rate of target tables. 

During re-ranking, we use a scoring function balancing multiple metrics for each table $t$ based on the hundreds of tables recalled in the coarse ranking stage. 
The overall calculation formula is as follows:
\begin{equation}
\text{Score}(t) = \text{Sim}(t) + \alpha \cdot \text{Embed}(t) + \beta \cdot \text{Heat}(t),
\label{eq:combined}
\end{equation}
where $\text{Embed}(t)$ represents the semantic recall score during the coarse ranking, which is the cosine similarity score of the vectorized results. 
$\text{Heat}(t)$ represents the table's popularity score, which is calculated by Tencent Data Warehouse based on query logs and execution task history.
$\text{Sim}(t)$ represents the similarity score between the user query and the table calculated from the field dimension, and $\alpha$, $\beta$ are hyperparameters. 

Regarding the token-level scores, user queries frequently reference specific keywords, while tables may contain thousands of fields.
Simple vectorizing the entire table's information may obscure some information details, leading to suboptimal recall performance. 
To address this shortcoming, we design a token-level similarity calculation to achieve better granular information recall.
Denote the i-th table as $t_{i}$, and the j-th column in $t_{i}$ as $c_{i,j}$. 
Assume that $t_{i}$ contains $N$ fields, and the input query contains $K$ keywords.
Then, the corresponding similarity score of $t_{i}$ is the sum of similarity calculations between the table and $K$ keywords.
The calculation formula is as follows, 
\begin{equation}
\text{Sim}(t_{i}) = \sum_{l=1}^{K} \text{sim}(k_{l},t_{i}),
\end{equation}
where $\text{sim}(k_{l},t_{i})$ represents the similarity score between keyword $k_{l}$ and table $t_{i}$, which is the maximum value of the text similarity with all columns:
\begin{equation}
\text{sim}(k_{l},t_{i}) = \max_{j \in \{ 1, 2, \ldots, N \}} \text{sim}(k_{l}, c_{i,j}).
\end{equation}

While TF-IDF~\cite{joachims1997probabilistic/tfidf} remains effective for real-time matching on short text queries, we  perform coarse filtering by restricting candidate tables to those with $(\text{table count} \times \text{field count}) \leq k$ dynamically adjusted per query complexity.  

\subsubsection{Dynamic Strategy Switching Driven by Data Conditions}
In practical BI applications, the data conditions of different clients vary significantly: some clients have abundant labeled data, while others can only provide a limited number of labeled samples due to frequent business changes or data privacy restrictions. Similarly, existing SQL generation methods are often tailored to specific scenarios or domains, making migration between private scenarios costly in terms of both deployment effort and data labeling. To address this challenge, we propose a \textit{data-condition-driven dynamic strategy switching mechanism}. This mechanism adaptively selects either a one-stage (high-precision fine-tuning) or two-stage (semantic inter-
mediate representation) generation strategy based on two key factors: the volume of labeled data and the semantic similarity between domains. By doing so, it effectively balances accuracy and cost-efficiency.

Formally, we define the number of available labeled Query-SQL pairs as the Labeled Data Volume($N_{labeled}$). We also quantify the semantic similarity between the target domain and an existing labeled domain (referred to as the source domain) as Domain Similarity($S_{domain}$)This similarity is computed via the cosine similarity between the embeddings of schema keywords provided by the business side, as follows:
$$
 \mathrm{Embed}(\text{target})=\frac{1}{|\phi_{\text{target}}|} \sum_{k \in \phi_{\text{target}}} \mathrm{Embed}(k),
$$
$$
\mathrm{Embed}(\text{source})=\frac{1}{|\phi_{\text{source}}|} \sum_{k \in \phi_{\text{source}}} \mathrm{Embed}(k),
$$
\begin{equation}
 \label{eq:domain_similarity}
 S_{\text{domain}} = \mathrm{cosine} \biggl(\mathrm{Embed}(\text{target}), \mathrm{Embed}(\text{source})\biggr),
\end{equation}
where $\phi_{target}$ and $\phi_{source}$ represent the sets of keywords for the target domain and the source domain, respectively (e.g., in the financial domain: "revenue", "quarterly report"). $ \text {Embed}( \cdot )$ is Sentence-BERT~\cite{sentence-bert} encoding function.

Based on the aforementioned features, we use rule-based criteria to determine the SQL generation strategy. Specifically:
\begin{align}
\label{eq:decision_rule}
\text{Strategy} = \begin{cases}
    \text{One-Step}, & \text{if } N_{\text{labeled}} \geq \alpha \ \land \ S_{\text{domain}} \geq \beta \\
    \text{Two-Step},  & \text{if } N_{\text{labeled}} < \alpha \ \lor \ S_{\text{domain}} < \beta 
\end{cases}
\end{align}
where $\alpha$ and $\beta$ are hyperparameters that are typically specified based on the actual situation. In our practical experience, $\alpha$ is usually set to 500 and $\beta$ is set to 0.7.

\subsubsection{One-Step SQL Generation}

\begin{figure}[htbp!]
    \centering
    \begin{subfigure}[b]{\linewidth}
        \centering
        \includegraphics[width=\linewidth]{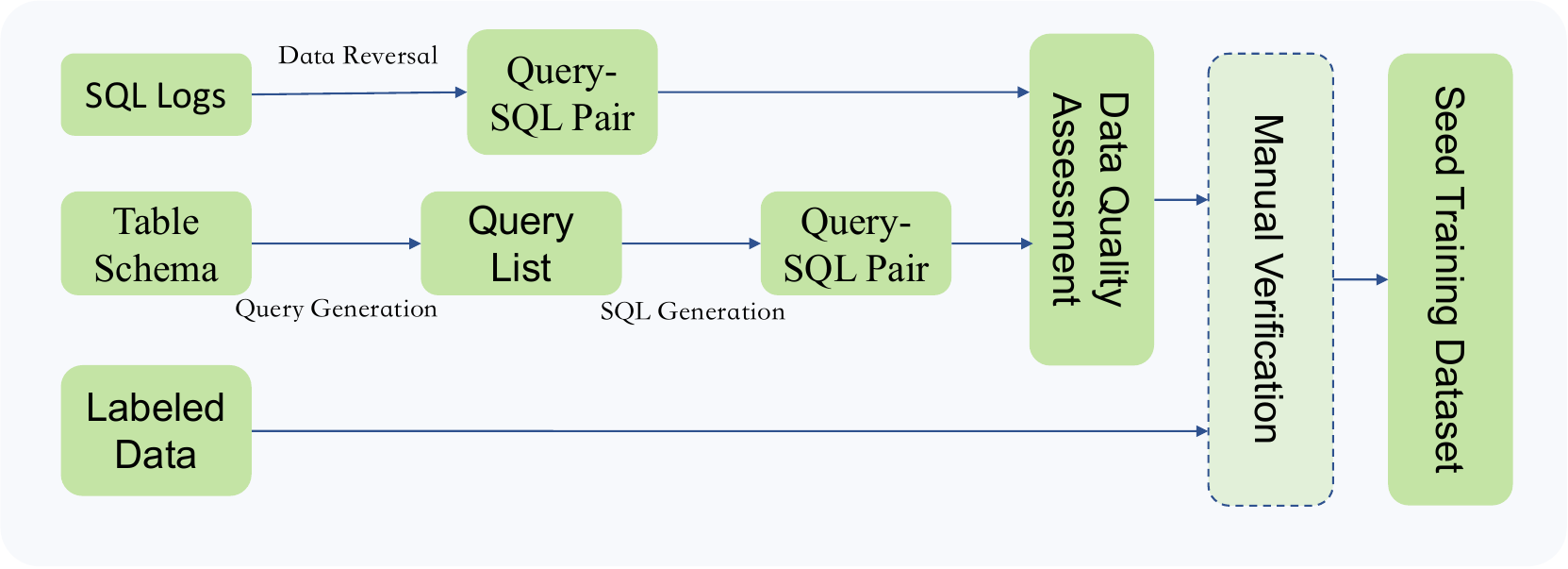}
        \caption{Data Generation}
        \label{subfig:data_gen}
    \end{subfigure}
    \begin{subfigure}[b]{\linewidth} 
        \centering
        \includegraphics[width=\linewidth]{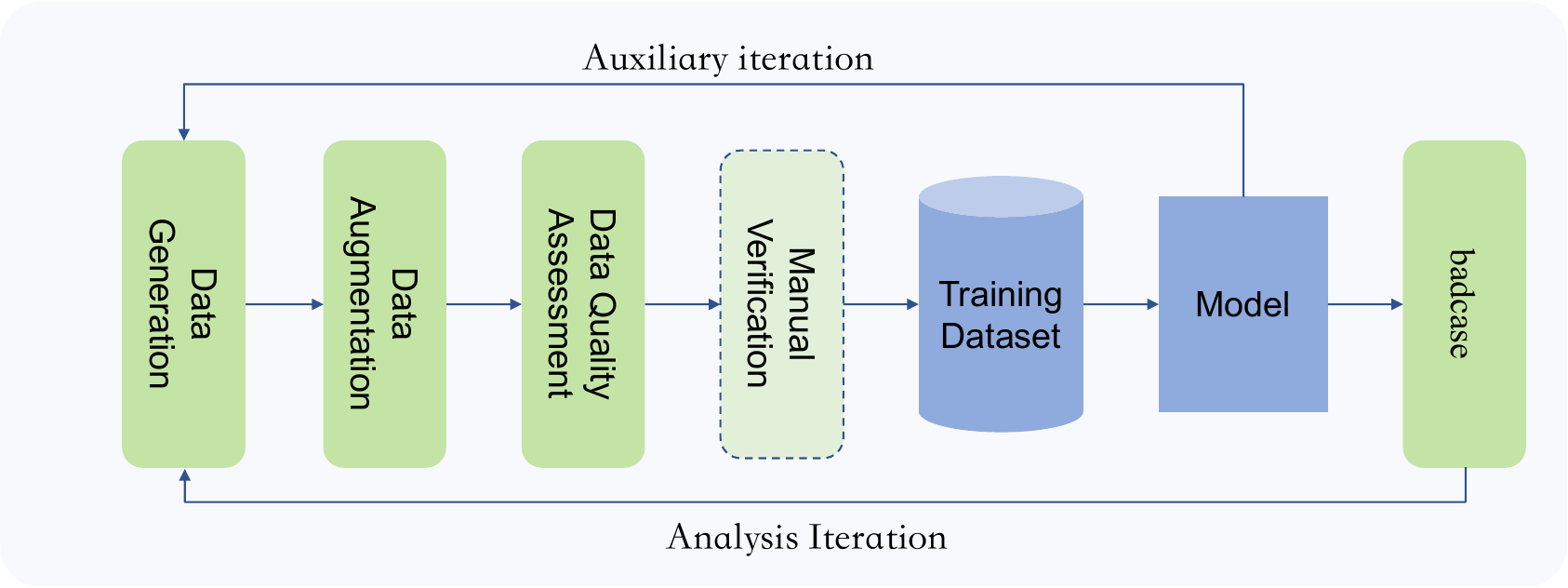}
        \caption{Automate Data Preparation Pipeline}
        \label{subfig:data_pipeline}
    \end{subfigure}
    \vspace{-4mm} 
    \caption{
         (a) illustrates the workflow for data generation, while  (b) presents the overall data preparation pipeline. In this pipeline, manual verification is an optional module.
    }
    \label{fig:data_pre}
    \vspace{-5pt}
\end{figure}

To improve the performance of supervised fine-tuning and reduce annotation efforts, we have developed an automated and iterative data preparation workflow guided by manually labeled data and identified bad cases. The overall pipeline is illustrated in Figure~\ref{fig:data_pre}, consisting of two main modules: data generation and data augmentation.

\textbf{Data Generation.}
To perform data generation, we need to utilize the SQL logs from the relevant scenarios and the corresponding schema information of the databases. SQL logs are the actual SQL code executed by users, representing the most authentic application data. However, since these are directly coded by users in the compiler, we need to use data reverse engineering to capture the users' true intentions.
The data reverse engineering functionality is primarily implemented through LLM-based prompts. The input information includes SQL, the corresponding schema information, and proprietary domain knowledge base information. On the other hand, we generate some question sets using the schema combined with LLM, and then construct some query-SQL pairs through a general SQL generation model.
The query-SQL pairs obtained through these two methods will enter the data quality evaluation module. This module mainly uses the locally deployed Qwen2.5-72B~\cite{qwen2.5} model for filtering, combined with manual evaluation to select the final high-quality training set. The manually annotated data will serve as the In-Context Learning~\cite{dong2022survey/in-context} (ICL) demonstrations used in the aforementioned process. The specific data generation pipeline can be seen in Figure ~\ref{subfig:data_gen}.

\textbf{Data augmentation}.
This stage enhances data quality through two key approaches: diversity expansion and error-based negative example injection. First, we use Self-Instruct~\cite{self-instruct} with prompts to generate a broader variety of SQL queries from seed data, significantly increasing the dataset's size and diversity. These queries are then evaluated by a LLM~\cite{zhujiu}, and only high-quality examples are selected for training. Second, we introduce negative examples by randomly rewriting cases from the enhanced dataset, based on error categories provided by the client. From experience, negative examples typically account for 5\% of the training set, helping improve the model’s robustness by teaching it to recognize and avoid common mistakes. Together, these strategies ensure a diverse, challenging dataset that enhances the model’s accuracy and reliability.

Once the data generation and augmentation phases are complete, the refined dataset is used to train the model. If necessary, the pipeline can be iterated to further refine the dataset based on additional error analysis or user feedback, ensuring continuous improvement in data quality. This iterative refinement continues until the dataset meets the desired standards or until the project reaches the budget constraints specified by our clients.

\textbf{Comparison with the previous work on data augmentation for NL2SQL systems}.
Early efforts in this field~\cite{wu2021data, weir2020dbpal} were generally simple, often relying on basic templates covering only a single table, heuristic methods such as synonym replacement, or traditional deep learning models like RNNs to construct NL-SQL pairs. These approaches greatly limited the quality of the generated data. 
More recent work, such as ScienceBenchmark~\cite{zhang2023sciencebenchmark}, is closer to our method, proposing an automatic data augmentation system to expand training data. The main differences are:
(1) \textit{Seed Data Source}: We use real SQL logs combined with data reverse engineering and NL-to-SQL translation to obtain Query-SQL pairs, while ScienceBenchmark relies on manually created data.
(2) \textit{Augmentation Goals}: Besides increasing data diversity, we focus on improving model robustness by injecting noise and negative examples; ScienceBenchmark mainly aims to enlarge dataset size to formalize a high-quality benchmark.
(3) \textit{Augmentation Approach}: We directly prompt LLMs to rewrite Query-SQL pairs, whereas ScienceBenchmark uses a multi-step pipeline of template extraction, population, NL question generation, and candidate selection.

\subsubsection{Two-Step SQL Generation}

To address the need for cross-domain generalization in real-world scenarios and the frequent occurrence of cold-start problems in business applications, we propose a novel two-stage SQL generation paradigm. This paradigm divides the SQL generation process into two stages: Semantic Intermediate Representation (SIR) and SQL Generation.

In the first stage, the construction of SIR focuses on abstracting user intent and data semantics, transforming complex natural language queries into a standardized intermediate form that is easy to handle. 
The rewriting process primarily relies on in-context learning~\cite{dong2022survey/in-context} to dynamically reconstruct user inputs. Additionally, we introduce a Chain-of-Thought (CoT)~\cite{wei2022chain/cot} strategy to explicitly guide the model in step-by-step reasoning about the latent intent and constraints of the query, thereby enhancing the quality of the rewritten results. 

Take the query \textit{``Analyze consumption of Tencent Cloud BI in 2023''} as an example. We first extract domain-specific knowledge which includes explanations of terms (\textit{``Cloud BI''} in this case) and top-$k$ demonstrations exhibiting highest similarities with the embeddings of the query as few-shot~\cite{fewshot} references. The input information in the SIR phase includes field descriptions retrieved from the knowledge base as well as few-shot demonstrations. The output of the SIR phase is shown below:
\begin{lstlisting}[
  frame=single,           % 单线边框
  basicstyle=\small\ttfamily, % 字体大小和等宽字体
  breaklines=true,        % 自动换行
  captionpos=b,           % 标题位置（如果需要）
  showstringspaces=false  % 隐藏字符串中的空格标记
]
{
    "Key Components": 
    {
        "Product Classification": "'Tencent Cloud BI' corresponds to product Chinese name (p_name_zh), specifically 'Tencent Cloud BI-Advanced Edition'",
        "Metric": "consumption_income_tax_inclusive (consumeincome)",
        "Time Range": 2023
    },
    "Knowledge Mapping": [
        "User's 'Tencent Cloud BI' refers to product Chinese name (p_name_zh)",
        "Only 'Tencent Cloud BI-Advanced Edition' is required for filtering"
    ],
    "Query Understanding": "User requires total tax-inclusive consumption income of 'Tencent Cloud BI-Advanced Edition' products in 2023 grouped by product name and year.", 
    "Rewritten Query": "Query the total consumption income (tax included) ( consumeincome ) of the product whose product name ( p_name_zh ) is 'Tencent Cloud BI-Advanced Edition' in 2023."
}
\end{lstlisting}

In the second stage, we use a code-rich LLM for SQL generation. The \textit{"Rewriting Query"} and the schema information corresponding to the user query together form the prompt. By first constructing and refining user queries before generating SQL, this two-step approach ensures high accuracy while enabling seamless migration across diverse industrial scenarios.
\section{Experiments and Results}
To evaluate our framework, in this section, we conduct an in-depth investigation from two insights:
1) On general NL2SQL tasks, \sys outperforms the state-of-the-art baselines on both SRD and MRD datasets; 
2) The end-to-end design of \sys is rational, where the individual performance of each module demonstrates significant superiority.

\subsection{Experiment Setups}

\subsubsection{Datasets}
To demonstrate the overall performance of \sys and the effectiveness of its various modules, we conduct comprehensive experiments using both academic benchmarks and industrial-level datasets on multiple tasks. We have released our carefully curated academic benchmark, MRD-BIRD. Due to data privacy concerns, we provide partially anonymized versions of two industrial-level datasets. They are publicly available at \url{https://github.com/TencentBigData-SiriusAI/SiriusBI}.
Detailed descriptions of the datasets used for each task are provided below:

\textit{\textbf{SRD NL2SQL.}}
To validate the effectiveness of the SQL generation module, we utilize the \textit{BIRD}~\cite{DBLP:conf/nips/LiHQYLLWQGHZ0LC23/bird-benchpaper}, a challenging large-scale database NL2SQL evaluation benchmark aimed at narrowing the gap between academic research and practical applications. It includes 95 large databases and high-quality (user question, SQL) pairs covering 37 professional fields.
Besides, we collect 182 SRD queries spanning 10669 tables and involving 43 columns from 5 different business scenarios, thereby creating a SRD dataset termed as \textit{SRD-Industry}.

\textit{\textbf{MRD NL2SQL.}}
This task better simulates real-world BI scenarios by evaluating both intent understanding and SQL generation in multi-round dialogue (MRD) settings. Due to the lack of a standard MRD NL2SQL benchmark, we expand part of the SRD dialogues from the BIRD dataset into 3–5 round dialogues. Specifically, we use an LLM to decompose complex SQL into turn-level queries, then generate corresponding user questions that reflect natural MRD conversation style—building on prior context without repetition. 
We also create \textit{MRD-Industry}, a new dataset matching the database scale of SRD-Industry, with 55 query sets spanning 2–5 rounds each (160 queries over 35 columns). To increase complexity, we select 10 sets and remove certain metrics and dimensions, and select another 10 sets in which we inject confusion by replacing metrics or dimensions with ambiguous alternatives.

\textit{\textbf{Knowledge Management.}}
To evaluate the knowledge management module, we focus on two main aspects: 1) Assessing its impact on NL2SQL tasks; 2) Evaluating the performance of knowledge retrieval. 
In the latter evaluation, we utilize the knowledge base corresponding to the MRD-Industry dataset, which contains knowledge across six dimensions: table, column, value, term, udf, and alias, totaling 7704 pieces of knowledge. 

\textit{\textbf{Data Insight.}}
For evaluation of data insights, we select 35 descriptive requests and 45 complex requests involving descriptive analysis used in industrial scenarios. 
The requests involve data retrieval, extreme value calculation, correlation analysis, etc.
Note that, other types of requests are not included in the dataset, as it's difficult to evaluate the correctness of answers to open-ended questions.

\subsubsection{Evaluation Metrics}
For NL2SQL tasks on the BIRD dataset, we adhere to the standard evaluation metrics as outlined in \cite{DBLP:conf/nips/LiHQYLLWQGHZ0LC23/bird-benchpaper}, utilizing \textit{Execution Accuracy} (EX) and \textit{Valid Efficiency Score} (VES). 
For NL2SQL tasks on our internal datasets, we follow previous works \cite{DBLP:journals/corr/abs-2405-00527/chatbi, floratou2024nl2sql} and employ the \textit{Useful Execution Accuracy} (UEX) metric, which considers a query correct if its intent aligns with that of the golden query. This makes UEX more suitable for business intelligence scenarios. 

\subsubsection{Baselines and Model Configurations}
We compare the NL2SQL capability of \sys with three state-of-the-art baselines as below:
(1) \textit{DIN-SQL}~\cite{DBLP:conf/nips/PourrezaR23/dinsql};
(2) \textit{MAC-SQL}~\cite{DBLP:journals/corr/abs-2312-11242/macsql};
(3) \textit{MRD-SQL}~\cite{DBLP:journals/corr/abs-2405-00527/chatbi};
(4) \textit{\sys}.
For the baseline models, we followed their publicly available usage instructions as closely as possible, including base models and prompt designs. For DIN-SQL and MAC-SQL, we used their original configurations and prompts without modification. We reproduce the NL2BI module of the previous work~\cite{DBLP:journals/corr/abs-2405-00527/chatbi} as MRD-SQL.
In our method and MRD-SQL, we use Qwen2.5-Coder-32B~\cite{hui2024qwen2coder} as the base model for two-step SQL generation on SRD and MRD datasets. For one-step SQL generation on the BIRD dataset (and MRD-BIRD), we use XiYanSQL-QwenCoder-32B~\cite{xiyansql}. This choice is motivated by its demonstrated state-of-the-art performance on this benchmark, which obviates the need for additional fine-tuning.
For industrial datasets, we fine-tuned Qwen2.5-Coder-32B to better fit domain data. Specifically, we use four nodes equipped with a total of 32 NVIDIA A100-40GB GPUs, setting the training to 3 epochs, a learning rate of \textit{3e-5}, a batch size of 2, and a sequence length of 8192 tokens. 
The fine-tuning dataset comprises over 9,000 entries collected from real-world industrial scenarios.
Overall, we ensured fair comparison by aligning baseline setups with their original methods and optimizing prompts within their intended use.

\subsection{End-to-End NL2SQL Performance}
In this experiment, we evaluate the end-to-end NL2SQL performance of various baselines on the SRD and MRD datasets. 
Note that, as an analysis module following NL2SQL, the data insight module is not included in this part.

\subsubsection{Results on SRD Datasets}

We conduct experiments on two SRD datasets—BIRD’s dev set and SRD-Industry—with results summarized in Table~\ref{tab:srd-result}. Our method, \sys, employing both one-step and two-step SQL generation, consistently outperforms baseline approaches. The improvement is especially pronounced on the SRD-Industry dataset, where \sys achieves at least a 40\% increase in UEX. This significant gain is attributed to \sys’s explicit design to tackle real-world challenges such as ambiguous fields and wide tables, which general-purpose models often struggle with. We exclude MRD-SQL from this comparison since it differs from \sys only in the multi-round dialogue analysis module, leading to identical performance on SRD datasets. Additionally, the one-step variant of \sys, fine-tuned on domain-specific data, surpasses the two-step method by 4.05\% in UEX. This suggests that supervised fine-tuning provides a measurable advantage in handling domain-specific queries, further enhancing its performance.

\begin{table}[htbp!]
  \centering
  \vspace{-10pt}
  \caption{Evaluations on the BIRD's dev set and SRD-Industry for SRD NL2SQL tasks.}
  \vspace{-10pt}
  \begin{tabular}{l|cc|c}
    \toprule
    \multirow{2}{*}{\makecell{\textbf{Methods}}} & \multicolumn{2}{c|}{\textbf{BIRD}} &  \textbf{SRD-Industry} \\
    & \textbf{EX(\%)} & \textbf{VES(\%)} & \textbf{UEX(\%)} \\
    \hline
    DIN-SQL + GPT-4o & 50.72 & 58.79 & 30.06 \\
    MAC-SQL + GPT-4o & 57.56 & 58.04 & 36.42 \\
    \sys \textit{two-step} & \underline{65.32} & \underline{67.87} & \underline{79.19} \\
    \sys \textit{one-step} & \textbf{68.97} & \textbf{70.89} & \textbf{83.24} \\
    \bottomrule
  \end{tabular}
  \vspace{-10pt}
  \label{tab:srd-result}
\end{table}

\subsubsection{Results on MRD Datasets}
Next, we evaluate \sys on two MRD datasets: MRD-BIRD and MRD-Industry, and present the experimental results in Table~\ref{tab:mrd-results}. Our findings reveal that NL2SQL methods designed for single-round dialogue perform poorly on MRD datasets. Specifically, the UEX scores of DIN-SQL and MAC-SQL on the MRD-Industry dataset fall below 16\%, highlighting their limitations in handling multi-round dialogue scenarios. In contrast, the two-step and one-step variants of \sys achieve UEX scores of 55.00\% and 62.50\%, respectively, demonstrating the consistent superiority compared to MRD-SQL in real-world NL2SQL tasks.

\begin{table}[htbp!]
    \centering
    \vspace{-5pt}
    \caption{Evaluations on the MRD-BIRD and MRD-Industry for MRD NL2SQL tasks.}
    \vspace{-10pt}
    \begin{tabular}{l|c|c}
    \toprule
    \textbf{Methods} & \makecell[c]{\textbf{MRD-BIRD} \\ \textbf{EX(\%)}} &  \makecell[c]{\textbf{MRD-Industry} \\ \textbf{UEX(\%)}} \\
    \hline
    DIN-SQL + GPT-4o & 41.69 & 11.25 \\
    MAC-SQL + GPT-4o & 42.35 & 15.63 \\
    MRD-SQL \textit{two-step} & 46.98 & 22.50  \\
    MRD-SQL \textit{one-step} & \underline{50.61} & 28.13  \\
    \sys \textit{two-step} & 48.03 & \underline{55.00}  \\
    \sys \textit{one-step} & \textbf{51.14} & \textbf{62.50}  \\  
    \bottomrule
    \end{tabular}
    \label{tab:mrd-results}
    \vspace{-5pt}
\end{table}

It is worth noting that on the MRD-BIRD dataset, the performance gap between \sys and MRD-SQL is narrower. Specifically, the one-step variant of MRD-SQL achieves a higher EX score than the two-step variant of \sys. We attribute this phenomenon to the absence of user interaction for intent clarification in the experiments on MRD-BIRD, which diminishes the advantage of \sys. However, by incorporating domain-specific knowledge during intent clarification, \sys still outperforms MRD-SQL when utilizing the same SQL generation method, underscoring the effectiveness of our approach in leveraging domain-specific insights to enhance performance.

\subsubsection{In-production Results}
In addition to experiments on industrial datasets, we evaluate \sys in production across multiple Tencent business domains. Before deployment, we apply the proposed data-condition-driven dynamic strategy switching mechanism to select the most suitable SQL generation approach based on each domain’s data characteristics.
For Tencent Finance and Advertising, clients provide sufficient labeled data ($N_{\text{labeled}} \geq \alpha$) and the source and target domains are identical, resulting in a semantic similarity $S_{\text{domain}}=1 \geq \beta$. Under these conditions, the mechanism selects the one-step fine-tuning strategy, achieving in-production accuracies of 97\% and 93\%, respectively.
Conversely, Tencent Cloud serves diverse users with limited labeled data and low domain similarity between our available labeled data. Thus, the mechanism chooses the two-step strategy, which attains an average accuracy of 96\% across 14 randomly selected clients.

\subsubsection{Discussion}
We observe that the in-production results are higher than those reported on the SRD and MRD datasets. This is mainly due to three factors: 1) Knowledge Completeness: Although (MRD-)BIRD datasets provide extra knowledge, their coverage is limited compared to our comprehensive internal knowledge base, which better supports accurate SQL generation in real scenarios. 2) Data Distribution and Query Complexity: Real-world user queries in specific applications tend to be more uniform and simpler, allowing fine-tuned models to achieve higher accuracy. In contrast, SRD and MRD contain diverse and complex questions from multiple domains, making them more challenging benchmarks. 3) Iterative Feedback and Model Refinement: In-production models benefit from multiple rounds of feedback and continuous improvement guided by real failure cases (see Section 3.4.3). SRD and MRD results, however, reflect single-cycle training without such iterative updates, limiting their performance. In summary, richer knowledge, simpler real-world queries, and ongoing model refinement explain the higher accuracy observed in production compared to benchmark datasets.

\subsection{Detailed Analysis}

\subsubsection{Evaluation of MRD-Q}
In this part, we present the ablation study on dialogue analysis and intention querying using the MRD-Industry dataset.
The experimental results are displayed in Table~\ref{tab:mrd_q}.
Based on the experimental results, we can observe that, both the \sys \textit{one-step} and \sys \textit{two-step} heavily rely on the dialogue analysis and the intention querying modules. 
Specifically, when both modules are removed (without MRD-Q), the performance of \sys declines significantly, with drops of 15.62\% and 34.38\% for one-step and two-step designs, respectively. 
Similarly, the absence of the intention querying module results in a performance degradation of 6.25\% and 21.25\% for \sys. 
This indicates that the dialogue analysis and intention querying modules are critical to the overall effectiveness of \sys. 

\begin{table}[htbp!]
    \centering
    \vspace{-5pt}
    \caption{Ablation study of MRD-Q on the MRD dataset. ``Q'' refers to the Intention Querying module.}
    \vspace{-10pt}
    \begin{tabular}{>{\centering\arraybackslash}p{4cm}>{\centering\arraybackslash}p{1cm}>{\centering\arraybackslash}p{1cm}}
    \toprule
    \textbf{Baseline} & \textbf{UEX} & \textbf{$\Delta$}  \\ \midrule
    \sys \textit{two-step w/o MRD-Q} & 20.62\% & -34.38\% \\
    \sys \textit{two-step w/o Q} & 33.75\% & -21.25\% \\
    \sys \textit{two-step } &  55.00\% & - \\
    \midrule
    \sys \textit{one-step w/o MRD-Q} & 46.88\% & -15.62\% \\
    \sys \textit{one-step w/o Q} & 56.25\% & -6.25\% \\
    \sys \textit{one-step} & 62.50\% & - \\  \bottomrule
    \end{tabular}
    \label{tab:mrd_q}
\vspace{-5pt}
\end{table}

In addition, the ablation study reveals that \sys \textit{one-step} fine-tuned on a private dataset exhibits greater robustness compared to \sys \textit{two-step}.
Concretely, the impact of removing intention querying in \sys \textit{one-step} is less pronounced, resulting in a performance drop of only 6.25\%, compared to a more significant decline of 21.25\% for \sys \textit{two-step}.
The difference can be attributed to the fact that \sys \textit{one-step} acquires domain-specific knowledge from the training data during the learning process, which helps to mitigate the negative effects of lacking conversational understanding and intention querying.

\subsubsection{Evaluation of Table Selection}
While table selection directly impacts the UEX of SQL selection, we conduct an ablation study on this component. 
We use the queries in the SRD dataset and evaluate the performance using the Recall@5 metric as described before. 
The experimental results are shown in Table ~\ref{tab:table-recall}.

\begin{table}[htbp!]
    \centering
    \vspace{-10pt}
    \caption{Ablation study of table selection. ``Embed'' and ``Heat'' refer to the embedding and heat information, respectively. }
    \vspace{-10pt}
    \label{tab:table-recall}
    \begin{tabular}{>{\centering\arraybackslash}p{3.5cm}>{\centering\arraybackslash}p{1.25cm}>{\centering\arraybackslash}p{1.25cm}}
    \toprule
    \textbf{Baseline} & \textbf{Recall@5} & \textbf{$\Delta$} \\ \midrule
    \sys \textit{w/o Embed \& Heat} & 77.47\% & -13.74\% \\
    \sys \textit{w/o Heat} & 79.12\% & -12.01\% \\
    \sys \textit{w/o Embed} & 83.52\% & -7.69\% \\
    \sys & \textbf{91.21\%} & - \\  \bottomrule
    \end{tabular}
\vspace{-10pt}
\end{table}

It is clear that both the embedding and heat information enhance the recall performance of table selection, which indicates the effectiveness of the combined ranking function in Equation~\ref{eq:combined}.

\subsubsection{Evaluation of Knowledge Management}
Then, we examine the impact of knowledge management (KM) on the NL2SQL pipeline.
The results of disabling this module on the MRD dataset are presented in Table ~\ref{tab:km-experiment}. 
The findings reveal a significant decline in performance for \sys when KM is turned off.
Concretely, UEX decreases by 23.12\% for \sys \textit{one-step} and 26.87\% for \sys \textit{two-step}.
These results highlight the critical role of the knowledge management module in enhancing the overall performance of NL2SQL, as both intention querying and SQL generation heavily depend on the availability of relevant knowledge.

Next, we evaluate the module using a dataset for knowledge retrieval with 7704 samples.
In this evaluation, we compare \sys against two widely used baseline models: BM25~\cite{robertson1976relevance/bm25} and Vector~\cite{DBLP:conf/emnlp/KarpukhinOMLWEC20/vector}, both of which are prominent in the field of text retrieval.
We assess the performance of these methods using two key metrics: KR Recall measuring the recall rate for retrieving relevant knowledge entries across all keywords and SL Recall measuring the recall rate specifically for linking table columns. 
The experimental results demonstrate the following performance: BM25 achieves a KR Recall of 90.28\% and an SL Recall of 89.58\%; Vector achieves a KR Recall of 93.06\% and an SL Recall of 93.06\%; \sys achieves a KR Recall of 95.14\% and an SL Recall of 93.75\%. \sys outperforms the baselines across both metrics, thereby ensuring that other modules leveraging knowledge management within \sys can access accurate and relevant additional information.

\begin{table}[htbp!]
    \centering
    \vspace{-5pt}
    \caption{Ablation study of knowledge management on MRD Dataset. ``KM'' refers to Knowledge Management.}
    \vspace{-10pt}
    \begin{tabular}{>{\centering\arraybackslash}p{4cm}>{\centering\arraybackslash}p{1cm}>{\centering\arraybackslash}p{1cm}}
    \toprule
    \textbf{Baseline} & \textbf{UEX} & \textbf{$\Delta$} \\ \midrule
    \sys \textit{two-step w/o KM} &  28.13\% & -26.87\%  \\
    \sys \textit{two-step } &  55.00\%  & - \\
    \midrule
    \sys \textit{one-step w/o KM} &  39.38\% & -23.12\% \\
    \sys \textit{one-step} &  62.50\% & - \\  \bottomrule
    \end{tabular}
    \label{tab:km-experiment}
\vspace{-10pt}
\end{table}

\subsubsection{Data Insight Evaluations}
Finally, we conduct experiments on the Data Insight module. 
Remind that the data insight dataset contains 80 quantifiable questions involving comprehensive analysis and complex task planning.

For evaluation, we use real-world queries from the financial domain and conduct comparative experiments with three powerful LLM baselines: GPT-4o~\cite{OpenAI}, Qwen2.5~\cite{hui2024qwen2coder} and Commnd R+~\cite{commandr+}. 
The accuracy of responses to these queries serves as the evaluation criterion, and to ensure the reliability of the experimental results, each query is tested three times. 

The results demonstrate that the Data Insight module of \sys achieves an accuracy of 95\%, significantly outperforming the baseline models. Specifically, Command-R+ achieves an accuracy of 75\%, Qwen2.5 achieves 80\%, and GPT-4o achieves 83\%. These findings indicate that the Data Insight module of \sys maintains a notable performance advantage even against the most effective LLMs currently in use. This superior performance can be attributed to the function call architecture in \sys, which enables it to outperform GPT-4o by 12\% in comprehensive data analysis tasks.

Specifically, for complex task planning, we compare the Data Insight module with a method that only performs tool invocation on the aforementioned high-quality queries. 
Compared to the simple tool invocation method, the MRD-agent architecture adopted by \sys shows a significant advantage, improving the accuracy from 70\% to 100\% on this dataset. 

\subsection{Case Study}

In this part, we use an exemplar to demonstrate the whole workflow through a BI interaction. At the beginning, the user raises a question -- \textit{What is the income of the Company A in 2024?} The Dialogue Analysis module evaluates the semantic integrity of this query. In fact, this query contains both a ``dimension'' (2024) and a ``metric'' (income), making it semantically complete. The query then proceeds to the second stage.

The Intention Querying module recalls relevant information from the knowledge base based on the input query. Specifically, a piece of retrieved knowledge is a triplet -- (company, by default, without special instructions, refers to Tencent, ``What is gross profit of the company in 2023?''), which provides an explanation for the term ``company''. 
The Intention Querying module further prompts the user for clarification as there still exists ambiguity within the query.
In this case, the word ``\textit{income}'' is ambiguous (with tax or not), the user will be asked whether it refers to the column ``\textit{shouldincome}'' or ``\textit{shouldincome\_after}''.

Afterwards, the SQL generation module selects the appropriate tables and translates the user query into an SQL query. The  generated SQL statement is \texttt{SELECT SUM(shouldincome\_after) AS total\_income FROM revenue\_by\_quarter WHERE YEAR \\ (ftime) = 2024 AND cname = Company A}.

\begin{figure}[htbp!]
    \centering
    \includegraphics[width=\linewidth]{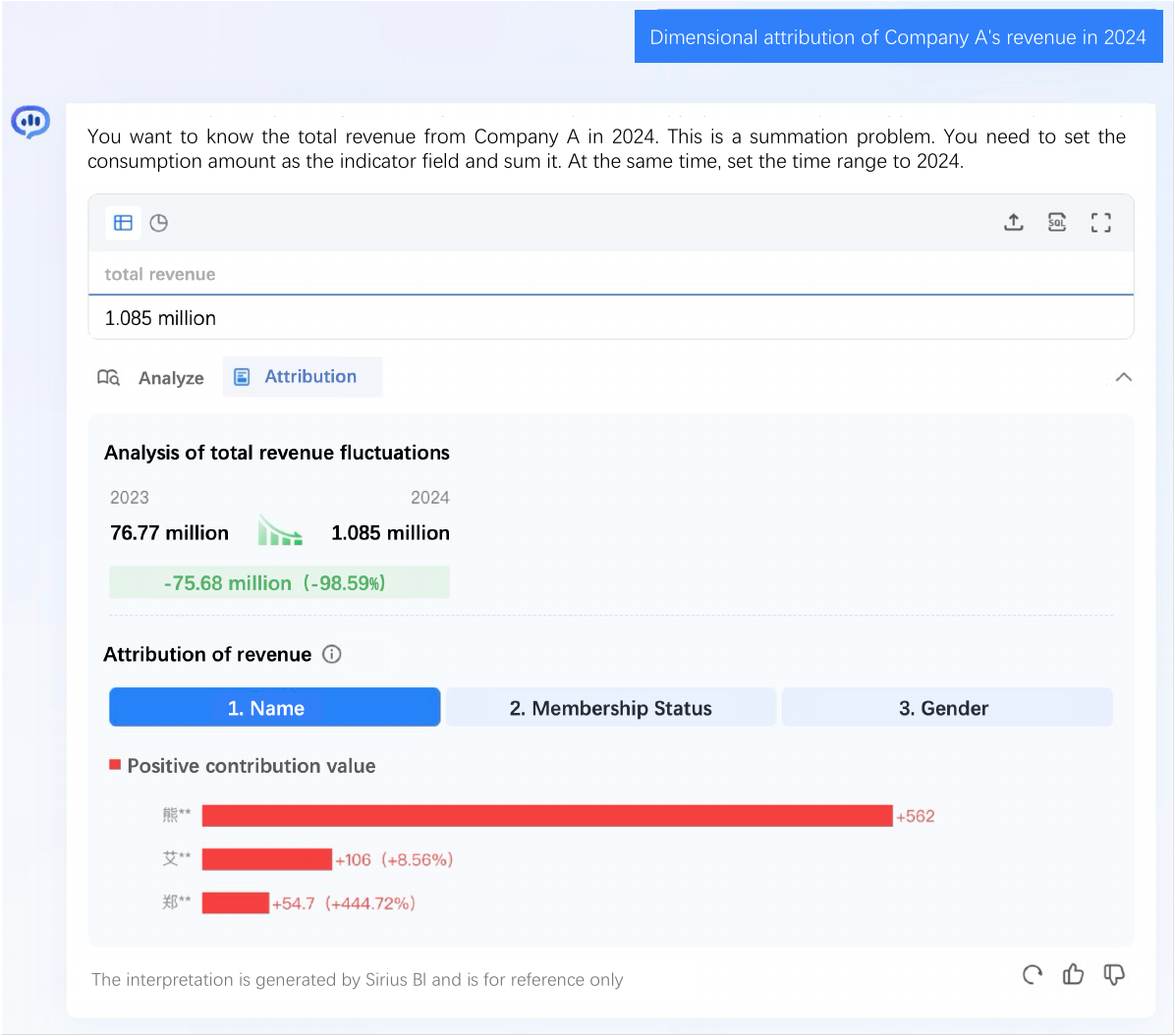}
    \vspace{-10pt}
    \caption{Showcase of practical application of dimensional attribution in data insights.}
    \label{insight_showcase}
    \vspace{-15pt}
\end{figure}

The Data Insight module receives the SQL queries and their corresponding execution results. 
Based on the user's additional insight requirements like \textit{``Perform dimension attribution analysis on the revenue of Company A in 2024''}, 
the module analyzes the user's intent, decomposes tasks through the Planner, performs data preparation and invokes attribution tools, and finally generates an insight report,
as shown in Figure ~\ref{insight_showcase}. 

\subsection{Utility and Usability Study}

To evaluate whether \sys effectively improves users' productivity and experience in real-world scenarios, we conducted a gray-release test within a business analysis team at Tencent data platform. This study aimed to quantify the system's impact on operational efficiency and gather subjective feedback on usability. We randomly divided 40 business analysts into two groups -- Group A (20 analysts): granted access to \sys for daily tasks; Group B (20 analysts): continued using legacy tools. The experiment lasted for two weeks, during which we monitored both groups’ workflows. To ensure comparability, we controlled that tasks were distributed evenly across groups based on complexity and historical completion times. We find that Group A exhibited a 30.2\% reduction in average task completion time compared to Group B. This efficiency gain is attributed to \sys's integrated workflow and advanced technical design. What's more, Group A provided an average SUS score of 84.2/100 (SD = 6.5), categorizing \sys as ``excellent'' according to the SUS benchmark~\cite{brooke1996sus}. Key strengths highlighted in feedback included: (1) 89\% of users reported that domain-specific knowledge grounding reduced manual schema lookup time. (2) 92\% agreed that \sys's data insight module ``provided reliable support for business decision-making'' (e.g., identifying root causes of revenue fluctuations). The gray-release test confirms that \sys enhances both productivity and user experience by addressing the functionality deficiencies.

\section{Conclusion}
In this paper, we propose \sys, a LLM-powered solution for business intelligence, specifically designed to address three key challenges encountered in industrial business intelligence scenarios.
Our system 1) establishes a comprehensive end-to-end workflow of the entire business intelligence process to address functionality deficiencies; 2) implements multi-round dialogue with querying capabilities to overcome interaction limitations; 3) adopts a data-conditioned SQL generation method selection strategy to alleviate the great cost brought by cross-domain migration in SQL generation.
Currently, \sys is deployed across Tencent's finance, advertising, and cloud sectors, serving enterprise clients with various profiles. 
Extensive experiments on public benchmark and industrial datasets demonstrate that \sys clearly outperforms competitive LLM-based baselines in both SRD and MRD-based NL2SQL tasks. 
Detailed analysis on different modules demonstrate the rationality of our design. 
In addition, ablation studies show that each component of \sys contributes significantly to the overall performance, indicating its robustness and scalability across domains.
User studies further confirm that \sys enhances both productivity and user experience.

\begin{acks}
This work is supported by National Natural Science Foundation of China (92470121, 62402016), National Key R\&D Program of China (2024YFA1014003), and High-performance Computing Platform of Peking University. 
We also thank the engineers from the Data Platform department, Technology and Engineering Group (TEG) of Tencent for their technical and engineering support.
The co-correspondence authors are Yang Li and Wentao Zhang.
\end{acks}

\balance

\bibliographystyle{ACM-Reference-Format}
\bibliography{siriusbi}  

\end{document}